\documentclass{article}
\usepackage{spconf,amsmath,graphicx}
\usepackage{cite}
\usepackage{multirow}
\usepackage{amsfonts}
\usepackage{array}
\usepackage{balance}

\title{Contextual colorization and denoising for low-light ultra high resolution sequences}
\name{N. Anantrasirichai and David Bull\thanks{This work was supported by Bristol+Bath Creative R+D under AHRC grant AH/S002936/1.}}
\address{Visual Information Laboratory, University of Bristol, UK}
\begin{document}
%

\maketitle

\begin{abstract}
Low-light image sequences generally suffer from spatio-temporal incoherent noise, flicker and blurring of moving objects. These artefacts significantly reduce visual quality and, in most cases, post-processing is needed in order to generate acceptable quality.  Most state-of-the-art enhancement methods based on machine learning require ground truth data but this is not usually available for naturally captured low light sequences. 
We tackle these problems with an unpaired-learning method that offers simultaneous colorization and denoising. Our approach is an adaptation of the CycleGAN structure. To overcome the  excessive memory limitations associated with ultra high resolution content, we propose a multiscale patch-based framework, capturing both local and contextual features. Additionally, an adaptive temporal smoothing technique is employed to remove flickering artefacts. Experimental results show that our method outperforms existing approaches in terms of subjective quality and that it is robust to variations in brightness levels and noise.

\end{abstract}
\section{Introduction}
\label{sec:intro}

\begin{figure}[t!]
	\centering
      		 \includegraphics[width=\columnwidth]{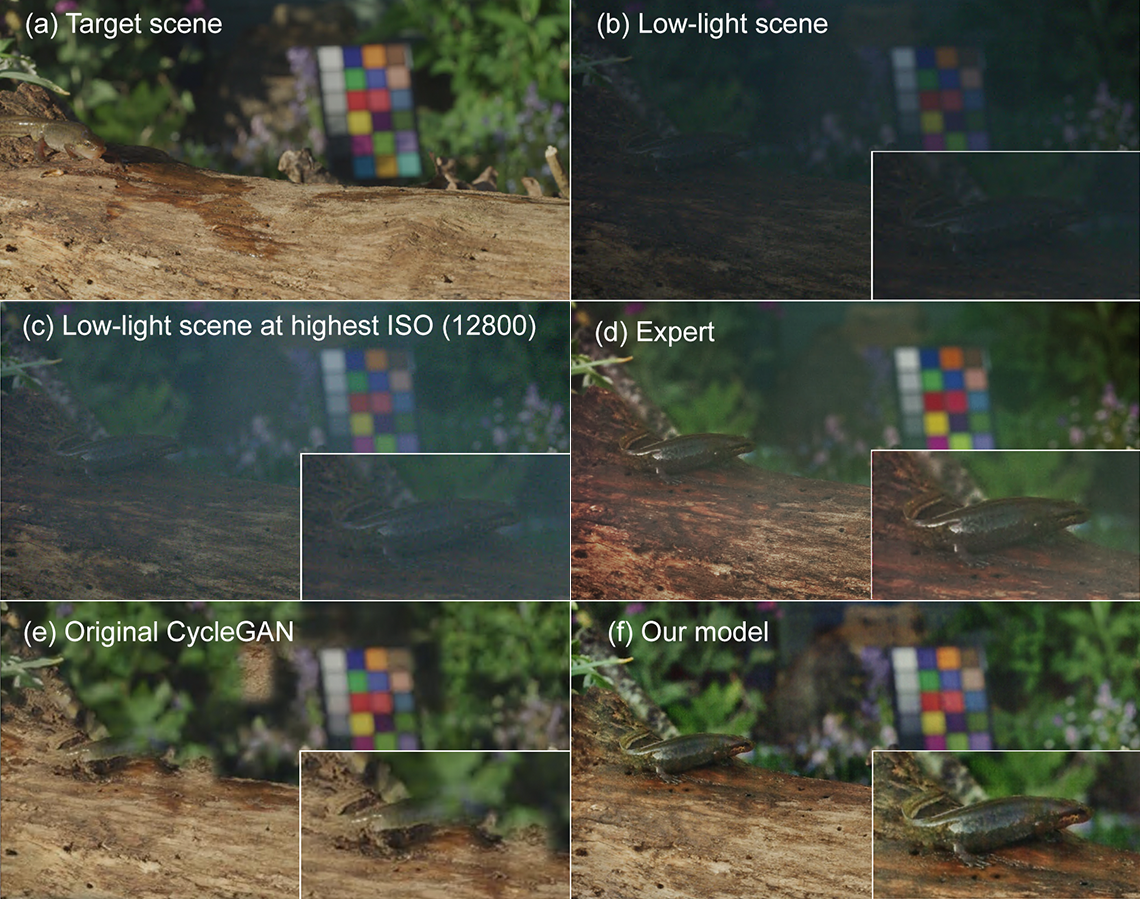}  
      		 \vspace{-7mm}
					\caption{\small (a-c) The 5K `\textit{Macro}' scenes. (d-f) enhancement results of the low-light scene (b) using (d)  manually editing by the expert, (e) CycleGAN \cite{Zhu:Unpaired:2017}, and (f) our model. Inset shows magnified object. }
    \label{fig:fig1_showoff}
\end{figure}

Low-light conditions can be problematic for video acquisition causing poor scene visibility (Fig. \ref{fig:fig1_showoff}b), focusing difficulties, blurring of moving objects due to limited of shutter speeds, and noise due to  high ISO values (Fig.~\ref{fig:fig1_showoff}c). These impairments are not only visually unpleasant, but they also impact upon the performance of automated tasks, such as classification, detection, and tracking. 

Traditional enhancement techniques typically wash out details, flatten appearance and amplify noise. In professional low-light applications, such as natural history filmmaking,  a specialist will manually apply colour grading and noise reduction techniques as part of the post-production workflow (e.g. Fig.~\ref{fig:fig1_showoff}d). The final results may however be unsatisfactory as the information contained in the source sequence is limited.

Recently, deep learning algorithms have demonstrated their effectiveness for image enhancement, segmentation, detection and denoising \cite{anantrasirichai:AI:2020}. Typical algorithms use  Convolutional Neural Networks (CNNs) to extract semantic meaning from low-level features, effectively working as an encoder. A convolutional decoder is then appended to produce a new image output \cite{Ronneberger:Unet:2015, Anantrasirichai:DefectNET:2019}. This module can be further employed as the generator in a Generative Adversarial Networks (GANs) \cite{Goodfellow:GAN:2014}, where a second module, the discriminator, is employed to improve the generator's performance by checking whether the received image is `real' or `fake'. Despite the success of these methods, their application to processing low light data is challenging due to the absence of ground truth data (replicating the same scene, registered at the pixel level, with appropriate lighting is practically impossible.

This paper presents a new end-to-end enhancement framework based on a generative model. It does not require paired training samples, but instead performs a mapping based on learnt common statistics. This approach, commonly referred to as a CycleGAN \cite{Zhu:Unpaired:2017}, does not manipulate the low-light input directly, but instead generates entirely new images. In our case, our aim is to transform noisy, low-light images (Fig. \ref{fig:fig1_showoff}b) into clean, sharp day-light versions (Fig. \ref{fig:fig1_showoff}a). It should be noted that an expert edited version could in principle be used as a target for the training process.  However, this process is hugely time consuming and expensive. 

Our framework is specifically tailored to ultra high resolution (UHR) sequences, which suffer from excessive memory requirements. To address this, we propose a patch-based strategy where a local patch is concatenated with the resized region where it belongs. The local patch contains localised features and noise characteristics, whilst the region patch contains contextual information. We hence calculate the training losses of the local and region patches separately, by using an $\ell_1$ loss for the local patches to minimise noise and preserve textures, and by using a perceptual loss function \cite{Ledig:Photo:2017} for the region patches to learn context. Finally, we propose an adaptive temporal smoothing technique to handle brightness changes and mitigate temporal inconsistencies. 

The remainder of this paper is organised as follows. A summary of related work is presented in Section \ref{sec:relatedwork}. Details of the proposed framework and our contributions are described in Section \ref{sec:method}. The performance of the method is evaluated in Section \ref{sec:results}, followed by the conclusions in Section \ref{sec:conclusion}.

\section{Related work}
\label{sec:relatedwork}

This section reviews state-of-the-art methods for image-to-image translation and denoising.  A survey of recent techniques can also be found in \cite{anantrasirichai:AI:2020}.

\textbf{Image-to-image translation} aims to produce a new image that has a different appearance to the input but with similar semantic content.  Early algorithms employed CNNs to perform tasks such as converting grayscale tones to natural colors \cite{Zhang:colorful:2016} or photographs to stylistic paints \cite{Gatys:Neural:2016}. Subsequently, conditional GANs, such as Pix2Pix \cite{Isola:Image:2017}, were proposed and these further extended the range of possible applications, including converting road maps to aerial photographs, or a sketch into a coloured object. These methods invariably exploit supervised learning, requiring a paired training dataset. CycleGAN \cite{Zhu:Unpaired:2017}, DualGAN \cite{Yi:DualGAN:2017}  and DiscoGAN \cite{Kim:Disco:2017} architectures were then proposed to overcome this limitation by training two GANs with two groups of unpaired images, mapping the characteristics of one group onto the other. More recently, unpaired GAN-based methods have been developed to produce diverse outputs from a single input \cite{Lee:DRIT:2020}. 

\textbf{Denoising} techniques are now, almost entirely, based on deep learning approaches.  For example, a residual noise map of an image can be estimated using a Denoising CNN (DnCNN) \cite{Zhang:dncnn:2017} while for  video, spatial and temporal networks are concatenated in \cite{Claus:ViDeNN:2019}. FFDNet \cite{Zhang:FFDNet:2018} works on reversibly downsampled subimages. VNLnet combines a non-local patch search module with DnCNN \cite{Davy:nonlocal:2019}. TOFlow \cite{Xue:Video:2019}  offers an end-to-end framework that performs motion analysis and video processing simultaneously. GANs have also been employed to estimate a noise distribution which is subsequently used to augment clean data for training CNN-based denoising networks (such as DnCNN) \cite{Chen:image:2018}. GANs also have been employed for denoising medical images \cite{Yang:Low:2018}, but they are not popular in the natural image domain due to the limited data resolution of current GANs.  

\section{Methodologies}
\label{sec:method}

\begin{figure}[t!]
	\centering
      		 \includegraphics[width=\columnwidth]{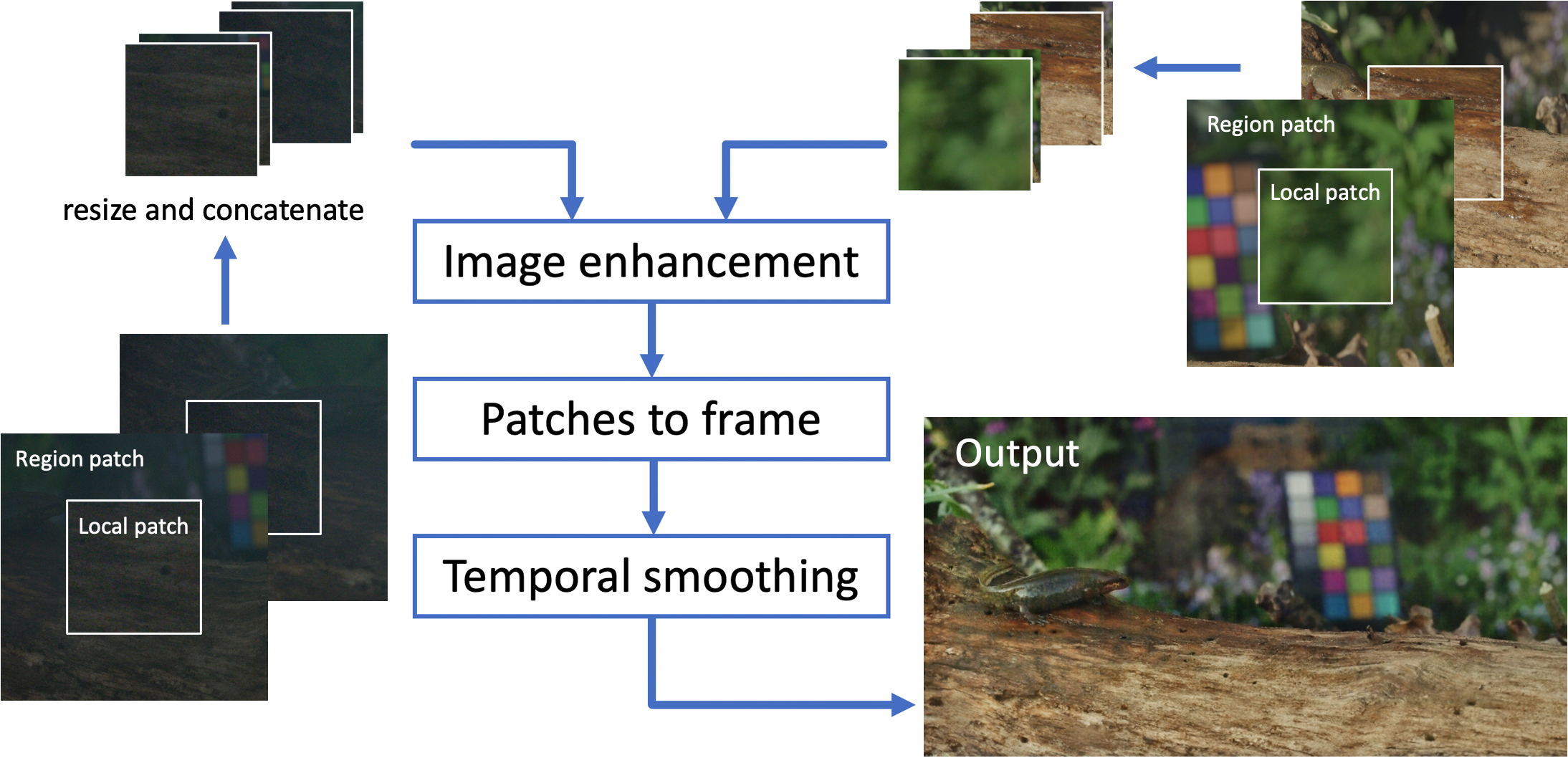}  
      		 \vspace{-7mm}
					\caption{\small Workflow. Left patches are low-light inputs and top right patches are targets.  }
    \label{fig:fig2_diagram}
\end{figure}

A diagram of the proposed framework is illustrated in Fig.~\ref{fig:fig2_diagram}. The process comprises patch generation, image enhancement, patch merging and temporal smoothing. 

\subsection{Patch generation}

Patches are cropped from the UHR images with the possible maximum size allowed by GPU memory. However, these may not contain sufficient contextual detail to be learnt by the CNNs. We therefore use both local and region- based patches. The sizes of the local and region patches are $N_l\times N_l$ and  $N_r\times N_r$ pixels, $N_l<N_r$, respectively. Region patches are cropped with the same centre point as the corresponding local patches, and then are resized to $N_l\times N_l$ pixels to concatenate with the local patches. The input to the GAN hence has six channels for an RGB colour format. The region size is not restricted, but it has to be large enough to capture the semantic meaning of the object in the local patch, e.g. close-up content requires larger region patches than landscape content.

\subsection{Enhancement with CycleGAN structure}
\label{ssec:network}

We model our image-to-image translation problem following the concept of CycleGAN \cite{Zhu:Unpaired:2017}. The first training group (group $A$) comprises the low-light patches and the second group (group $B$) comes from the target image. The patches of both groups are translated twice, i.e. from group A to group B with the generator $G_A$, then translated back to the original group A with the generator $G_B$. Then, the loss function compares the input image and its reconstruction. 

\textbf{Generators}: Both $G_A$ and $G_B$ have three sequential modules: i) an encoder with three convolutional blocks (3$\times$3 conv with strike=2 + instance norm + softshrink), ii) nine ResNet blocks \cite{He:ResNet:2016}, and iii) a decoder with three convolutional blocks (3$\times$3 conv + instance norm + ReLU). We choose ResNet over DenseNet  \cite{Huang:Densely:2017} and UNet \cite{Ronneberger:Unet:2015} because it requires less memory. As the low-light inputs are noisy, we use a learnable softshrink activation in the encoder. This is inspired by wavelet soft shrinkage \cite{Isogawa:deep:2018}. However we found that using softshirnk activation in all non-linear activation layers reduces micro contrast of the output, so we employ ReLU activations in the other two modules. 

\textbf{Discriminators}: Following the original CycleGAN, five convolutional blocks (4$\times$4 conv + instance norm + LeakyReLU slope=0.2) are used. We tested several deeper architectures but observed similar results at the cost of a higher memory requirement.

\textbf{Loss functions}: The training process aims to minimise a loss function $\mathcal{L}_\text{final} $ comprising i) adversarial loss $\mathcal{L}_\text{GAN}$, ii) cycle consistency loss $\mathcal{L}_\text{cyc}$, and iii) identity loss $\mathcal{L}_\text{idt}$, shown in Eq.\ref{eqn:finalloss}, where $\lambda_\text{GAN}$, $\lambda_\text{cyc}$ and $\lambda_\text{idt}$ are weights.
\begin{equation}
\label{eqn:finalloss}
\mathcal{L}_\text{final}  = \lambda_\text{GAN}\mathcal{L}_\text{GAN} + \lambda_\text{cyc}\mathcal{L}_\text{cyc} + \lambda_\text{idt}\mathcal{L}_\text{idt}.
\end{equation}
\noindent $\mathcal{L}_\text{GAN}$ joins a generator loss, to fool the discriminator,  and a discriminator loss, to distinguish between the real and translated samples. $\mathcal{L}_\text{cyc}$ enforces forward-backward consistency, and $\mathcal{L}_\text{idt}$ indicates that $G_A$ should be the identity if the target patch is fed and similarly with $G_B$ if the low-light patch is fed. We calculate the losses of the local patches ($A^l, B^l$) and region patches ($A^r, B^r$) separately, and weight toward the loss of the local patches. This is because the local patches are what we actually want to translate, whilst the region patches provide contextual information as guidance. That is, 
\begin{equation}
\mathcal{L}_{t} = w \mathcal{L}_{t}^{l} + (1-w) \mathcal{L}_{t}^{r}, w > 0.5,  t \in \{\text{GAN}, \text{cyc}, \text{idt}\}. 
\end{equation}
For $\mathcal{L}_\text{GAN}$, in addition to the original CycleGAN that uses least square GAN (LSGAN), we employ a relativistic average LSGAN (RaLSGAN) \cite{jolicoeur-martineau:relativistic:2019}, which measures the global probability of input data to be more realistic than the opposing type. This improves training stability and visual quality.

For $\mathcal{L}_\text{cyc}$ and $\mathcal{L}_\text{idt}$, we employ an $\ell_1$ loss for the local patches. This is a pixel-wise loss that is robust to noise and capable of preserving textures. For the region patches, we use a perceptual loss computed from feature maps $\phi$ extracted with a pretrained VGG19 \cite{Ledig:Photo:2017}. This has proven performance for measuring contextual similarity. We however employ  $\ell_1$-norm instead of  $\ell_2$-norm used in \cite{Ledig:Photo:2017} as it is more robust to outliers. $\mathcal{L}_\text{cyc}$ and $\mathcal{L}_\text{idt}$ are computed as follows.
\begin{subequations}
\label{eqn:cyclosses}
\begin{align}
\begin{split}
\mathcal{L}_\text{cyc}^l &= ||G_B(G_A(A^l)) - A^l||_1 \\
& \ \ \ + ||G_A(G_B(B^l)) - B^l||_1, 
\end{split}
\\
\begin{split}
\mathcal{L}_\text{cyc}^r &= ||\phi(G_B(G_A(A^r)))- \phi(A^r))||_1 \\
& \ \ \ + ||\phi(G_A(G_B(B^r))) - \phi(B^r)||_1,
\end{split}
\end{align}
\end{subequations}
\vspace{-3mm}
\begin{subequations}
\label{eqn:idtlosses}
\begin{align}
\mathcal{L}_\text{idt}^l &= ||G_A(B^l) - B^l||_1 + ||G_B(A^l) - A^l||_1, \\
\begin{split}
\mathcal{L}_\text{idt}^r &= ||\phi(G_A(B^r))- \phi(B^r))||_1 \\
& \ \ \ + ||\phi(G_B(A^r)) - \phi(A^r)||_1,
\end{split}
\end{align}
\end{subequations}

We also tried Hinge adversarial loss \cite{Zhang:self:2019}, style loss \cite{Gatys:Neural:2016}, wavelet-based loss, gradient loss \cite{Umer:Deep:2020} and total variation loss \cite{Yu:Generative:2018}. None of these improved enhancement performance.

\subsection{Patches to frame}
For inference, we divide each frame of the UHR sequence into overlapping patches. All results in this paper were reconstructed from the patches shifted by $N_l/2$ pixels. The input and the output of the network have six channels comprising the RGB local and RGB region patches, but only the RGB local patches are used to reconstruct a frame. The patches are merged with Gaussian weights ($\mu$=$N_l/2$, $\sigma$=$N_l/6$, where $\mu$ and $\sigma$ are the mean and the standard deviation).

\subsection{Temporal smoothing}
Due to memory limitations associated with UHR videos, learning process can only be performed on a frame-by-frame basis. Since this can lead to temporal inconsistency, a pixel-wise average across a temporal sliding window is used to smooth brightness and colour. The window size of each pixel is changed adaptively, based on the magnitude of its motion. Firstly, each frame in the sliding window is warped and registered to the current frame. We adapt a warping process using multi-scale gradient matching in \cite{Anantrasirichai:Atmospheric:2018} to reduce large displacements amongst frames, and then apply a wavelet-based registration \cite{Anantrasirichai:Atmospheric:2013} to mitigate micro misalignment. Motion estimation is performed using coarser level wavelet coefficients to determine large motion components and then finer level coefficients to refine the motion field.  The sliding window is defined at the pixel level with the maximum values of $N_\text{max}$ backward and $N_\text{max}$ forward frames. The pixels with larger motion will be constructed with a fewer frames, whilst the stable pixels will be the average of all 2$N_\text{max}$+1 in the sliding window. We set $N_\text{max}$ to 6 frames and if the motion is more than 256 pixels, no neighbouring frames are used.

\begin{figure*}[t!]
	\centering
      		 \includegraphics[width=\textwidth]{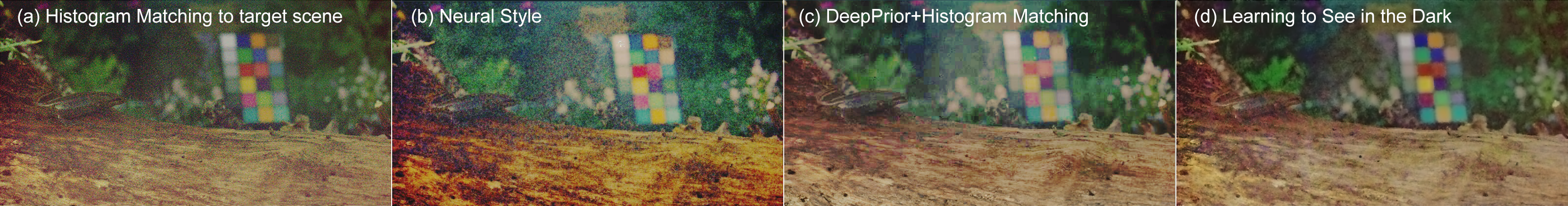}  
      		 \vspace{-7mm}
					\caption{\small Enhancement results of low-light `\textit{Macro}' sequence using (a) traditional histogram matching to the target, (b) Neural Style \cite{Gatys:Neural:2016}, (c) DeepPrior \cite{Lempitsky:Deep:2018} then histogram matching, and (d) Learning to see in the dark \cite{Chen:Learning:2018}.  }
    \label{fig:fig3_othermethods}
\end{figure*} 

\begin{figure*}[t!]
	\centering
      		 \includegraphics[width=\textwidth]{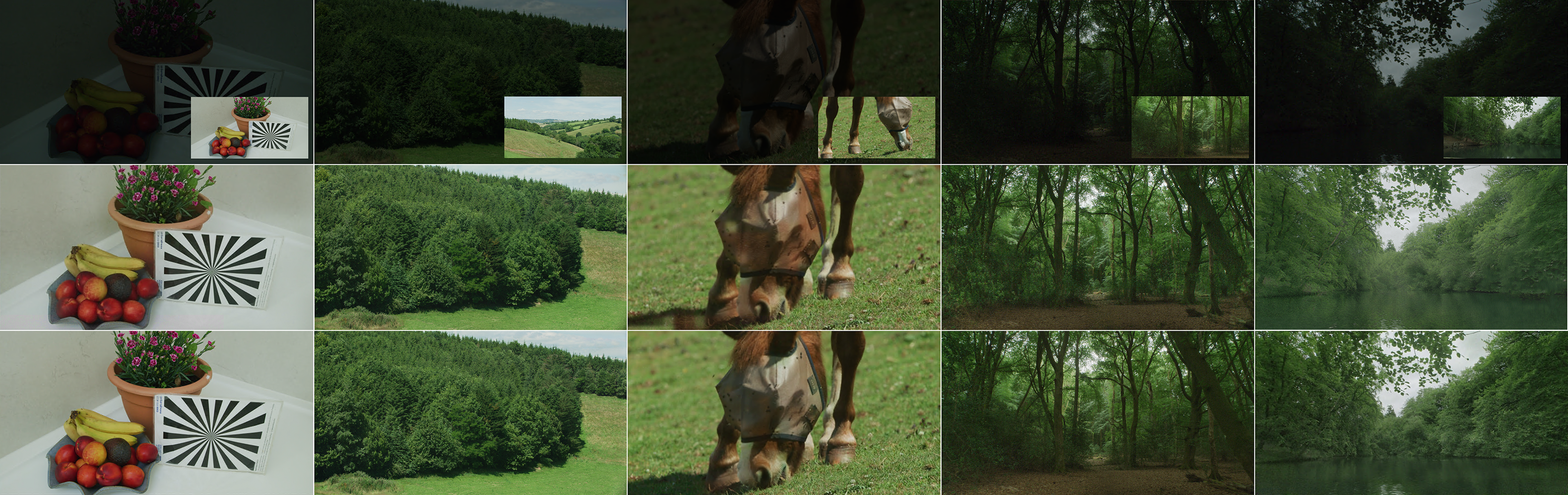}  
      		 \vspace{-7mm}
					\caption{\small Results of frame 200 of (left-right) `\textit{Static}', `\textit{Fly}', `\textit{Horse}' , `\textit{Woods}', and `\textit{River}. (Top-bottom) Low-light scene with the target scene in the inset, results of CycleGAN and our proposed method, respectively.}
    \label{fig:fig5_allresults}
\end{figure*} 

\begin{figure}[t!]
	\centering
      		 \includegraphics[width=\columnwidth]{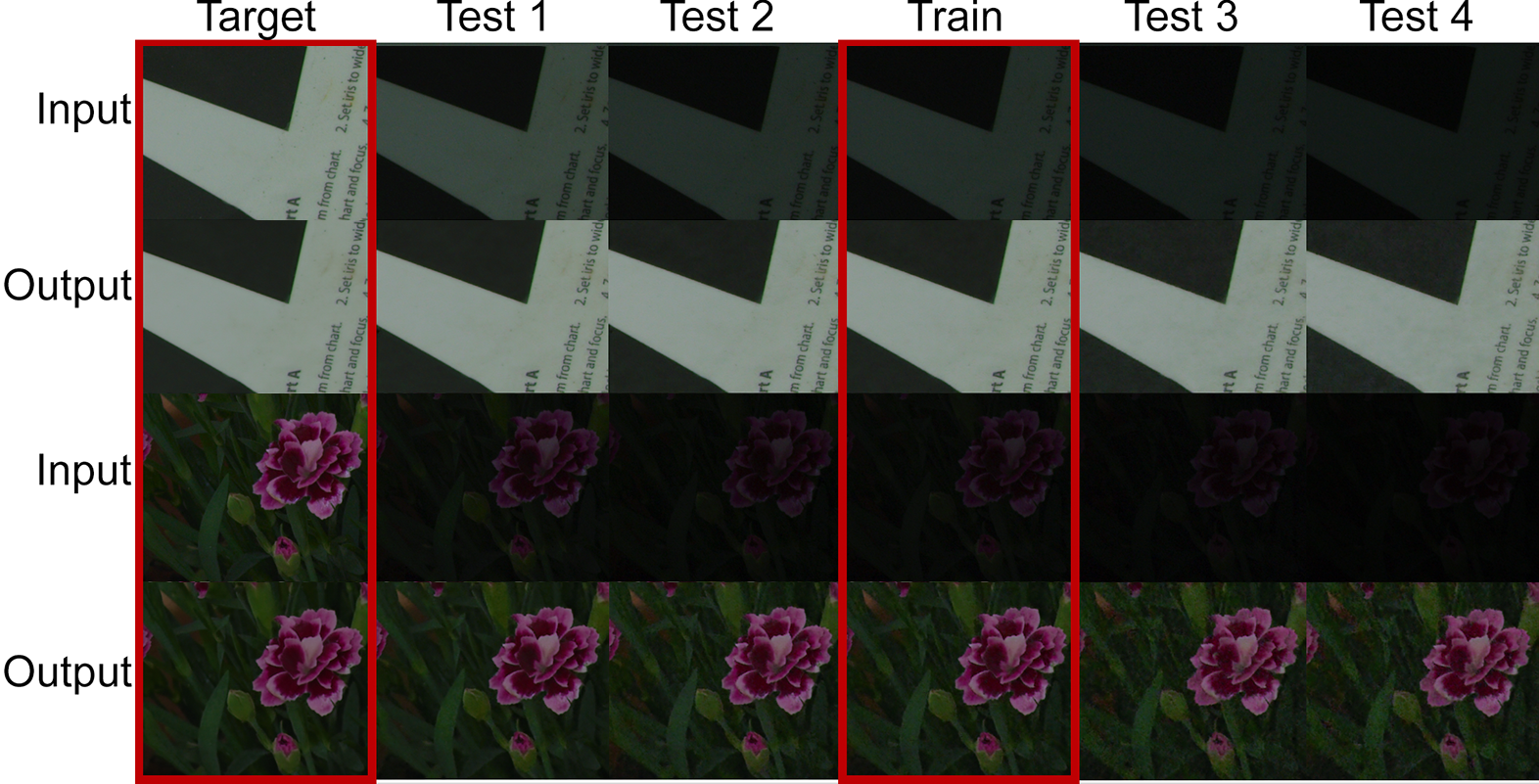}  
      		 \vspace{-7mm}
					\caption{\small Robustness test with various intensity changes of `\textit{Static}' sequence. The columns with the red blocks are training sets. The 1$^\text{st}$ and 3$^\text{rd}$ rows are inputs. The 2$^\text{nd}$ and 4$^\text{th}$ are outputs of our method.  }
    \label{fig:fig4_robust}
\end{figure} 

\section{Experiments and Discussion}
\label{sec:results}


The method was tested with six UHR sequences: i) three of 8K resolution (7680$\times$4320 pixels), named `\textit{Static}', `\textit{Fly}', and `\textit{Horse}', and ii) three of 5K resolution (5120$\times$2880 pixels), named `\textit{Macro}', `\textit{Woods}', and `\textit{River}'.  These were captured using RED Gemini cameras in R3D format, 25 fps (https://www.red.com/) and converted to TIF format. The low-light and target scenes were captured at different times and under different lighting. Their positions were not registered. The `\textit{Static}' sequence is static indoor scene, whilst the others contain moving objects and dynamic background. 

\textbf{Training parameters}: Local patches are memory limited to  360$\times$360 pixels and region patches are 1,000$\times$1,000 pixels and 1,500$\times$1,500 pixels for 5K and 8K sequences, respectively (apart from `\textit{Horse}' that is  2,500$\times$2,500 pixels). We randomly cropped 1,000 patches of the first frame of each sequence for training. Only the first frame was used as we wanted to investigate the robustness of the model.  Training parameters were: $\lambda_\text{GAN}$=1, $\lambda_\text{cyc}$=10, $\lambda_\text{idt}$=0.5, $w$=0.9.

\textbf{Results and comparison}: The results in Fig.~\ref{fig:fig1_showoff} and Fig.~\ref{fig:fig5_allresults} reveal that our method creates better contrast with lower noise than  CycleGAN, and that contextual information from region patches assists the formation of local information (e.g. the lizard head appears much more clearly in our result). We also provide comparisons with other automated methods: DeepPrior \cite{Lempitsky:Deep:2018} is an unsupervised denoising technique (needing subsequent histogram matching);  Learn-to-See-in-the-Dark is a supervised low-light image enhancement method \cite{Chen:Learning:2018}. We retrained their model with our `\textit{Static}' sequence combined with their datasets. The results in Fig. \ref{fig:fig3_othermethods} clearly show that residual noise remains a problem for these methods. 

\textbf{Robustness}:  Fig.~\ref{fig:fig4_robust} shows results for four brightness values: two between the training low-light and the target datasets, and two  darker values than the training version. The model can be seen to be robust to intensity changes. The effect of noise is noticeable in the darker sequences but is subtle, because the convolutional layers behave like low-pass filters. We also see that, as the input noise level reduces, the outputs become sharper and more vivid.

\vspace{-3mm}
\section{Conclusions}
\label{sec:conclusion}
\vspace{-3mm}
We present a novel end-to-end framework for joint colorization and denoising of low-light UHR sequences. Since registered ground truth is unavailable, we use a CycleGAN that learns statistics of the source and the target groups. To address the issue of memory load, we propose a patch-based technique, where local and region patches are concatenated as the input of the network.  The architecture of both generators and discriminators, as well as the loss functions, are modified to suit UHR images. Finally, we used an adaptive temporal smoothing technique to mitigate flickering artefacts. Our proposed framework clearly outperforms existing methods, providing evident benefits in terms of subjective quality.
\small

\vspace{-3mm}
\section{Acknowledgement}
\label{sec:Acknowledgement}
\vspace{-3mm}
We would like to thank Esprit film and television, and BBC Bristol for providing datasets.

\pagebreak
\balance
\small
\bibliographystyle{IEEEbib}
\bibliography{literature_review}

\end{document}